\documentclass{aa}
\usepackage{txfonts}
\usepackage{natbib}
\usepackage{graphicx}
\usepackage{url}

\def\ltsima{$\; \buildrel < \over \sim \;$}
\def\lsim{\lower.5ex\hbox{\ltsima}}

\bibpunct{(}{)}{;}{a}{}{,}

\begin{document}

\title{REM observations of GRB\,060418 and GRB\,060607A: the onset of the afterglow and the initial fireball Lorentz factor determination}

\author{E.~Molinari\inst{1}
\and S.D.~Vergani\inst{2,3}
\and D.~Malesani\inst{4,5}
\and S.~Covino\inst{1}
\and P.~D'Avanzo\inst{6,1}
\and G.~Chincarini\inst{7,1}
\and F.M.~Zerbi\inst{1}
\and L.A.~Antonelli\inst{8}
\and P.~Conconi\inst{1}
\and V.~Testa\inst{8}
\and G.~Tosti\inst{9}
\and F.~Vitali\inst{8}
\and F.~D'Alessio\inst{8}
\and G.~Malaspina\inst{1}
\and L.~Nicastro\inst{10}
\and E.~Palazzi\inst{10}
\and D.~Guetta\inst{8}
\and S.~Campana\inst{1}
\and P.~Goldoni\inst{11,12}
\and N.~Masetti\inst{10}
\and E.J.A.~Meurs\inst{2}
\and A.~Monfardini\inst{13}
\and L.~Norci\inst{3}
\and E.~Pian\inst{14}
\and S.~Piranomonte\inst{8}
\and D.~Rizzuto\inst{1,7}
\and M.~Stefanon\inst{15}
\and L.~Stella\inst{8}
\and G.~Tagliaferri\inst{1}
\and P.A.~Ward\inst{2}
\and G.~Ihle\inst{15}
\and L.~Gonzalez\inst{15}
\and A.~Pizarro\inst{15}
\and P.~Sinclair\inst{15}
\and J.~Valenzuela\inst{15}
}

\offprints{E. Molinari:\\ \email{emilio.molinari@brera.inaf.it}}

\institute{INAF - Osservatorio Astronomico di Brera, via E. Bianchi 46, I-23807 Merate (LC), Italy.
\and Dunsink Observatory - DIAS, Dunsink lane, Dublin 15, Ireland.
\and School of Physical Sciences and NCPST, Dublin City University - Dublin 9, Ireland.
\and International School for Advanced Studies (SISSA/ISAS), via Beirut 2-4, I-34014 Trieste, Italy.
\and Dark Cosmology Centre, Niels Bohr Institute, University of Copenhagen, Juliane Maries vej 30, DK--2100 K\o{}benhavn, Denmark.
\and Dipartimento di Fisica e Matematica, Universit\`a dell'Insubria, via Valleggio 11, I-22100 Como, Italy.
\and Universit\`a degli Studi di Milano Bicocca, piazza delle Scienze 3, I-20126 Milano, Italy.
\and INAF - Osservatorio Astronomico di Roma, via di Frascati 33, I-00040 Monteporzio Catone (Roma), Italy.
\and Dipartimento di Fisica e Osservatorio Astronomico, Universit\`a di Perugia, via A. Pascoli, I-06123 Perugia, Italy.
\and INAF-IASF di Bologna, via P. Gobetti 101, I-40129 Bologna, Italy.
\and APC, Laboratoire Astroparticule et Cosmologie, UMR 7164, 11 Place Marcelin Berthelot, F-75231 Paris Cedex 05, France.
\and CEA Saclay, DSM/DAPNIA/Service d'Astrophysique, F-91191, Gif-s\^ur-Yvette, France.
\and CNRS, Institut Néel, 25 rue des Martyrs, F-38042 Grenoble, France.
\and INAF - Osservatorio Astronomico di Trieste, via G.B. Tiepolo 11, I-34143 Trieste, Italy.
\and European Southern Observatory, Alonso de C\'ordova 3107, Vitacura, Casilla 19001, Santiago 19, Chile.
}

\date{Received  / Accepted }

\abstract
{Gamma-ray burst (GRB) emission is believed to originate in highly relativistic
fireballs.}
{Currently, only lower limits were securely set to the initial fireball Lorentz
factor $\Gamma_0$. We aim to provide a direct measure of $\Gamma_0$.}
{The early-time afterglow light curve carries information about $\Gamma_0$,
which determines the time of the afterglow peak. We have obtained early
observations of the near-infrared afterglows of GRB\,060418 and GRB\,060607A
with the REM robotic telescope.}
{For both events, the afterglow peak could be clearly singled out, allowing a
firm determination of the fireball Lorentz of $\Gamma_0\sim400$, fully
confirming the highly relativistic nature of GRB fireballs. The deceleration
radius was inferred to be $R_{\rm dec} \approx 10^{17}$~cm. This is much larger
than the internal shocks radius (believed to power the prompt emission), thus
providing further evidence for a different origin of the prompt and afterglow
stages of the GRB.}
{}


\titlerunning{The onset of the afterglow}
\authorrunning{Molinari et al.}

\maketitle

\section{Introduction}

The early stages of gamma-ray burst (GRB) afterglow light curves display a rich
variety of phenomena at all wavelengths and contain significant information
which may allow determining the physical properties of the emitting fireball.
The launch of the \textit{Swift} satellite \citep{Neil04}, combined with the
development of fast-slewing ground-based telescopes, has hugely improved the
sampling of early GRB afterglow light curves. To date there are several 
published early optical/near-infrared (NIR) afterglow light curves that cannot
be fitted with a simple power-law, testifying that the detections started
before the afterglow began its regular decay. To explain the early behaviour of
most of them (GRB\,990123: \citealt{Akerlof99}; GRB\,021004:
\citealt{KobZha03}; GRB\,021211: \citealt{Fox03,Li03}; GRB\,041219A:
\citealt{Fan05,Ves05}; GRB\,050730: \citealt{Pand06}; GRB\,050801:
\citealt{Rom06,Ryk06}; GRB\,050820A: \citealt{Ves06}; GRB\,050904:
\citealt{Boe06,Wei06}; GRB\,060206: \citealt{Wozn06,Monfardini06}; GRB\,060210:
\citealt{Sta07}) different mechanisms have been proposed, i.e. reverberation of
the prompt emission radiation, reverse and refreshed shocks and/or energy
injection and, lately, large angle emission \citep{PK07}.

Since many processes work in the early afterglow, it is often difficult to
model them well enough to be able to determine the fireball characteristics.
The simplest case is a light curve shaped by the forward shock only, as could
be the case for GRB\,030418 \citep{Ryk04} and GRB\,060124 \citep{Pat06}, but
the lack of a measured redshift in the former case and the poor sampling in the
latter prevented to derive firm conclusions. This case is particularly
interesting because, while the late-time light curve is independent of the
initial conditions (the self-similar solution), the time at which the afterglow
peaks depends on the original fireball Lorentz factor $\Gamma$, thus allowing a
direct measurement of this fundamental parameter \citep{SP99}. The short
variability timescales, coupled with the nonthermal GRB spectra, indeed imply
that the sources emitting GRBs have a highly relativistic motion
\citep{Ruderman75,Fenimore93,Piran00,Lith01}, to avoid suppression of the
high-energy photons due to pair production. This argument, however, can only
set a lower limit to the fireball Lorentz factor. Late-time measurements (weeks
to months after the GRB) have shown $\Gamma \sim \mbox{a few}$
\citep{Frail97,Taylor05}, but a direct measure of the initial value (when
$\Gamma$ is expected to be $\sim 100$ or more) is still lacking.

We present here the NIR early light curves of the GRB\,060418 and GRB\,060607A
afterglows observed with the REM robotic
telescope\footnote{\url{http://www.rem.inaf.it}} \citep{Zerb01,Chinc03} located
in La Silla (Chile). These afterglows show the onset of the afterglow and its
decay at NIR wavelengths as simply predicted by the fireball forward shock
model, without the presence of flares or other peculiar features.

\begin{figure}
\centering\includegraphics[width=0.82\columnwidth]{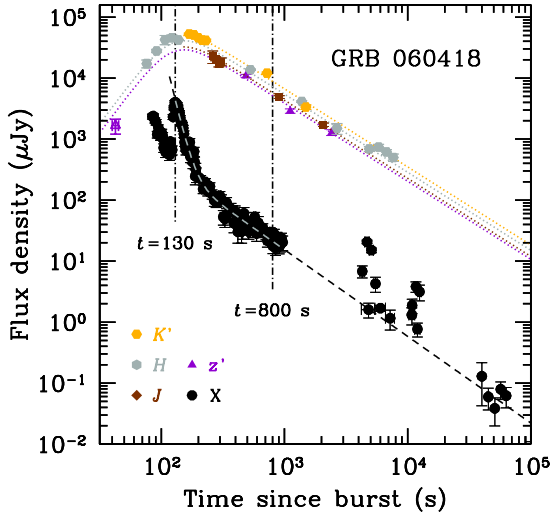}
\caption{NIR and X-ray light curves of GRB\,060418. The dotted lines show the
models of the NIR data using the smoothly broken power law (see Sect.
\ref{modelling}), while the dashed line shows the best-fit to the X-ray data.
The vertical lines mark the epochs at which we computed the SED
(Fig.~\ref{fg:SED}).\label{fig:060418}}
\end{figure}

\section{Data}

GRB\,060418 and GRB\,060607A were detected by \textit{Swift} at 03:06:08\,UT
\citep{Falc06a} and 05:12:13\,UT \citep{Ziaee06}, respectively. The BAT light
curve of the former ($T_{90} = 52 \pm 1$~s) showed three overlapping peaks
\citep{Cumm06}. For the latter, the light curve is dominated by a double-peaked
structure with a duration $T_{90} = 100 \pm 5$\,s \citep{Tuel06}. The
\textit{Swift} XRT started observing the fields 78 and 65~s after the trigger,
respectively. For both bursts a bright, previously uncatalogued fading source
was singled out. The XRT light curves show prominent flaring activity,
superposed to the regular afterglow decay. For GRB\,060418 a prominent peak,
also visible in the BAT data, was observed at about 128~s after the trigger
\citep{Falc06b} while the XRT light curve of GRB\,060607A was characterised by
at least three flares \citep{Page06}. UVOT promptly detected bright optical
counterparts for both events. The redshift is $z=1.489$ for GRB\,060418
\citep{Dupr06,Vrees06} and $z=3.082$ for GRB\,060607A \citep{Led06}.

The REM telescope reacted promptly to both GCN alerts and began observing the
field of GRB\,060418 64\,s after the burst (39\,s after the reception of the
alert) and the field of GRB\,060607A 59\,s after the burst (41\,s after the
reception of the alert). REM hosts REMIR, an infrared imaging camera that
operates in the range 1.0--2.3~$\mu$m ($z'JHK'$), and ROSS, an optical imager
and slitless spectrograph. For both targets a bright NIR source was identified
\citep{Cov06a,Cov06b}. ROSS could not observe these two GRBs due to maintenance
work. REM followed the two events down to the sensitivity limits adopting two
different observing strategies. In the former case multi-colour $z'JHK$-band
observations were carried out while in the latter case only the $H$ filter was
used in order to get a denser sampling of the light curve in just one band. All
our photometric data are reported in Tables~\ref{data1} and \ref{data2}.

\begin{figure}
\centering\includegraphics[width=0.82\columnwidth]{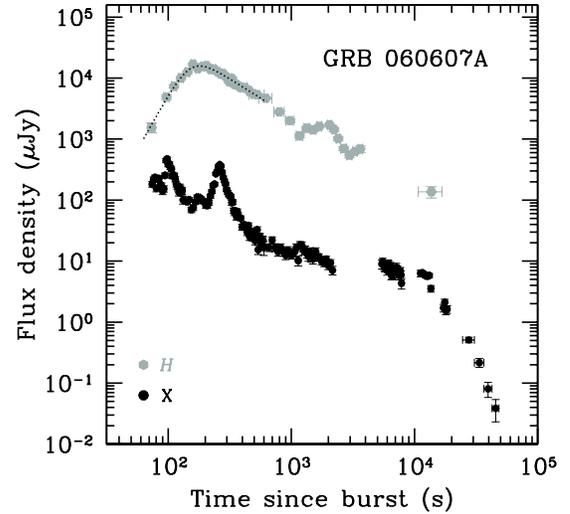}
\caption{$H$-band and X-ray light curves of GRB\,060607A. The dotted line shows
the model to the NIR data using the smoothly broken power law (see Sect.
\ref{modelling}).\label{fig:060607}}
\end{figure}

\section{Results and Discussion}

\subsection{Light curve modelling}
\label{modelling}

\begin{table}
\caption{Best fit values of the light curves of the first hour of observations
for GRB\,060418 and the first 1000\,s for GRB\,060607A (1$\sigma$ errors),
using the smoothly broken power-law reported in Sect. \ref{modelling} with
$t_{\rm peak} = t_{\rm b} (-\alpha_{\rm r}/\alpha_{\rm
d})^{1/[\kappa(\alpha_{\rm d}-\alpha_{\rm r})]}$. The relatively large $\chi^2$
of the fit results from small-scale irregularities present throughout the light
curve (see Figs. \ref{fig:060418} and \ref{fig:060607}).}
\centering
\begin{tabular*}{\columnwidth}{@{\extracolsep{\fill}}l@{\hfill}c@{\hfill}c@{\hfill}c@{\hfill}c@{\hfill}c@{\hfill}c}
\hline
GRB     & $t_{\rm peak}$ (s) & $t_{\rm b}$   (s)     & $\alpha_{\rm r}$      & $\alpha_{\rm d}$       & $\kappa$            & $\chi^2/{\rm d.o.f.}$ \\  \hline
060418  & $153_{-10}^{+10}$ & $127_{-21}^{+18}$  & $-2.7_{-1.7}^{+1.0}$  & $1.28_{-0.05}^{+0.05}$ & $1.0_{-0.4}^{+0.4}$ & $33.3/16$             \\
060607A & $180_{-6}^{+5}$   & $153_{-12}^{+12}$  & $-3.6_{-1.1}^{+0.8}$  & $1.27_{-0.11}^{+0.16}$ & $1.3_{-1.1}^{+0.9}$ & $28.5/19$             \\ 
\hline
\end{tabular*}

\label{tab:fit}
\end{table}

Figures~\ref{fig:060418} (GRB\,060418) and \ref{fig:060607} (GRB\,060607A) show
the NIR and X-ray light curves of the two afterglows. The X-ray data have
been taken with the \textit{Swift} XRT. The analysis was carried out adopting
the standard pipeline. Pile-up corrections both in WT and PC mode were taken
into account when required. The NIR light curves of the two events show a
remarkable similarity. Both present an initial sharp rise, peaking at
100--200~s after the burst. The NIR flux of GRB\,060418 decays afterwards as a
regular power law. The NIR light curve of GRB\,060607A shows a similar, smooth
behaviour up to $\sim 1000$~s after the trigger, followed by a rebrightening
lasting $\sim 2000$~s. 

To quantitatively evaluate the peak time, we fitted the NIR light curves using
a smoothly broken power-law \citep{Beu99}: $F(t) = {F_0}/ {\left[ (t/t_{\rm
 b})^{\kappa\alpha_{\rm r}} + (t/t_{\rm b})^{\kappa\alpha_{\rm d}}
 \right]^{1/\kappa}}$,
where $F_0$ is a normalisation constant, $\alpha_{\rm r(d)}$ is the slope of
the rise (decay) phase ($\alpha_{\rm r} < 0$) and $\kappa$ is a smoothness
parameter. The time at which the curve reaches its maximum is $t_{\rm peak} =
t_{\rm b} (-\alpha_{\rm r}/\alpha_{\rm d})^{1/[\kappa(\alpha_{\rm
d}-\alpha_{\rm r})]}$. We obtain for GRB\,060418 and GRB\,060607A peak times of
$153 \pm 10$ and $180 \pm 6$~s, respectively. Any other suitable functional
forms provides comparable results. The complete set of fit results is reported
in Table~\ref{tab:fit}. 

As for many other GRBs observed by \textit{Swift}, the early X-ray light curves
of both events show several, intense flares superimposed on the power-law decay
\citep{Chinca07}. In particular, for GRB\,060418 a bright flare was active
between $\sim 115$ and 185~s. Excluding flaring times, the decay is then
described by a power law with decay slope $\alpha_{\rm X} = 1.42 \pm 0.03$. The
X-ray light curve of GRB\,060607A is more complex and presents two large flares
within the first 400~s. After that the flux density decreases following a
shallow power law (with small-scale variability), until steepening sharply at
$t \sim 10^4$~s.

\begin{figure}
\centering\includegraphics[width=0.82\columnwidth]{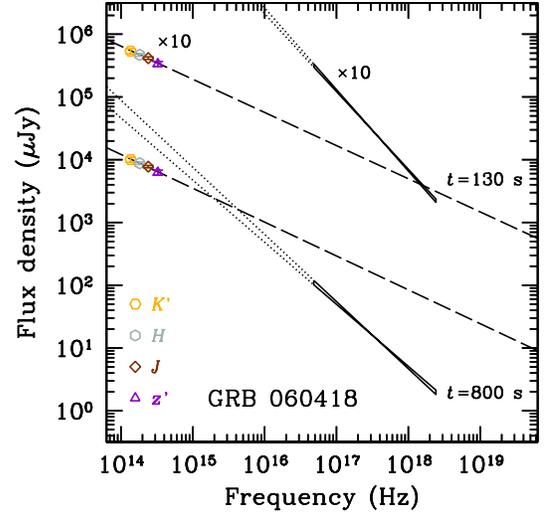}
\caption{SED of the GRB\,060418 afterglow at 130 and 800~s after the trigger.
NIR data have been corrected for Galactic and host extinction ($A_V = 0.74$ and
0.10 mag, respectively).\label{fg:SED}}
\end{figure}

By comparing the X-ray and NIR light curves of both bursts, it is apparent that
the flaring activity, if any, is much weaker at NIR frequencies. It is thus
likely that the afterglow peak, visible in the NIR, is hidden in the X-ray
region. In the case of GRB\,060418, thanks to our multicolour data, this can be
confirmed by considering the NIR/X-ray spectral energy distribution (SED;
Fig.~\ref{fg:SED}). At $t = 800$~s, when both the NIR and X-ray light curves
are decaying regularly, the SED is described by a synchrotron spectrum, with
the cooling frequency lying at $\nu_{\rm cool} \sim 3 \times 10^{15}$~Hz. The
spectral slopes at NIR and X-ray wavelengths are $\beta_{\rm NIR} = 0.65 \pm
0.06$ and $\beta_{\rm X} = 1.03 \pm 0.04$, respectively. A small amount of
extinction in the host ($A_V = 0.10$~mag assuming the SMC extinction curve) was
considered to make $\beta_{\rm NIR}$ consistent with the synchrotron
expectation $\beta_{\rm NIR} = \beta_{\rm X} - 0.5$. The situation is clearly
different at early times, when X-ray flaring is active: the X-ray component is
softer ($\beta_{\rm X} = 1.27 \pm 0.03$) and much brighter than the
extrapolation of the NIR emission (even including the dust correction). This
implies a different origin for the X-ray emission. For GRB\,060607A, the light
curve is complicated by several flares. As we lack multicolour data, a detailed
analysis is not possible. 

\subsection{Determination of the Lorentz factor $\Gamma_0$}

Our spectral and temporal analysis agrees with the interpretation of the NIR
afterglow light curves as corresponding to the afterglow onset, as predicted by
the fireball forward shock model \citep{SP99,Meszaros}. According to the fits,
the light curves of the two afterglows peak at a time $t_{\rm peak} > T_{90}$
as expected in the impulsive regime outflow (`thin shell' case). In this
scenario the quantity $t_{\rm peak}/(1+z)$ corresponds to the deceleration
timescale $t_{\rm dec} \sim R_{\rm dec}/(2c\Gamma_{\rm dec}^2)$, where $R_{\rm
dec}$ is the deceleration radius, $c$ is the speed of light and $\Gamma_{\rm
dec}$ is the fireball Lorentz factor at $t_{\rm dec}$. It is therefore possible
to estimate $\Gamma_{\rm dec}$ \citep{SP99}, which is expected to be half of
the initial value $\Gamma_0$ \citep{PK00,Meszaros}. For a homogeneous
surrounding medium with particle density $n$, we have
\begin{equation}
\Gamma(t_{\rm peak}) = \left[ \frac{3E_{\gamma}(1+z)^3}{32\pi n m_{\rm p} c^5 
\eta t_{\rm peak}^3} \right]^{1/8}\approx 160 \left[\frac{E_{\gamma,53}(1+z)^3}{\eta_{0.2}n_0 {t}_{\rm peak,2}^3}\right]^{1/8},
\label{Gamma}
\end{equation}
where $E_\gamma = 10^{53} E_{\gamma,53}$~erg is the isotropic-equivalent 
energy released by the GRB in gamma rays, $n = n_0$~cm$^{-3}$, $t_{\rm peak,2}
= t_{\rm peak}/(100~{\rm s})$, $\eta = 0.2\,\eta_{0.2}$ is the radiative
efficiency and $m_{\rm p}$ is the proton mass. We use
$E_{\gamma}=9\times10^{52}$\,erg for GRB\,060418 \citep{Golen06} and
$E_{\gamma}\sim1.1\times10^{53}$\,erg for GRB\,060607A \citep{Tuel06}.
Substituting the measured quantities and normalising to the typical values $n =
1$~cm$^{-3}$ and $\eta = 0.2$ \citep{Bloom03}, we infer for both bursts 
$\Gamma_0 \approx 400\,(\eta_{0.2}n_0)^{-1/8}$. This value is very weakly
dependent on the unknown parameters $n$ and $\eta$, and therefore provides a
robust determination of $\Gamma_0$.

One possible caveat concerns the density profile of the surrounding medium,
which is still uncertain. \cite{ChevaLi00} proposed a wind-shaped profile $n(r)
= Ar^{-2}$, where $A$ is a constant, as expected around a massive star. In
principle, the temporal and spectral properties of the afterglow can be used to
distinguish among the two cases. For $\nu < \nu_{\rm cool}$, before $t_{\rm
dec}$, the flux is expected to rise as $t^3$ for the homogeneous (ISM) case,
while it evolves slower than $t^{1/3}$ in a wind environment. The measured
values for GRB\,060418 and GRB\,060607A seem therefore consistent with the ISM
case (see also \citealt{Jin07}). For GRB\,060418, however, no closure relation
is satisfied for $t > t_{\rm peak}$. In fact, irrespective of the density
profile, the temporal and spectral slopes in the X-ray region (which lies above
$\nu_{\rm cool}$) should verify $\alpha_{\rm X} = 3\beta_{\rm X}/2 - 1/2$,
which is clearly not the case given the observed values. Such discrepancy has
been observed in many other examples and may be due to radiative losses, or
varying equipartition parameters, or Compton losses (see e.g.
\citealt{Pana05,zhang06}). A detailed treatment of these effects is beyond the
scope of this paper. We however provide an estimate of $\Gamma_0$ in the case
of a wind environment:
\begin{equation}
\Gamma(t_{\rm peak}) = \left[ \frac{E_{\gamma}(1+z)}{8\pi A m_{\rm p} c^3 
\eta t_{\rm peak}} \right]^{1/4}.
\end{equation}
Using the same values as above for $E$ and $z$, and assuming $A^* =
A/(3\times10^{35}~\mbox{cm}^{-1}) = 1$, we find $\Gamma_0 \approx 150$ for both
bursts.

The determination of $t_{\rm peak}$ is in principle affected by the choice of
the time origin $t_0$, which we set to the BAT trigger time
\citep{Lazzati06,Quim06}. These authors have however
shown that this effect is small, and mostly affects the rise and decay slopes rather than the peak time (especially when $t_{\rm peak}$ is larger than the burst duration, as
for GRB\,060418 and GRB\,060607A). The measurement of $\Gamma_0$ is thus not
very sensitive to the exact choice of $t_0$.

For both bursts, we could not detect any reverse shock emission. The lack of
such flashes has already been noticed previously in a set of \textit{Swift}
bursts with prompt UVOT observations \citep{Rom06}. Among the many possible
mechanisms to explain the lack of this component, strong suppression (or even
total lack) of reverse shock emission is naturally expected if the outflow is
Poynting-flux dominated \citep{Fan04,ZhangKob05}. Nevertheless, \cite{Jin07}
showed that for GRB\,060418 and GRB\,060607A the reverse shock emission might
be too weak to be detected.

Finally, the optical afterglow of GRB\,030418 showed a peaked light curve
similar to our cases, and has been explained by \cite{Ryk04} as due to
decreasing extinction. Our peaks are too sharp to match this interpretation.
Furthermore, a very large extinction would be implied, which contrasts with the
UVOT detection in the rest-frame UV \citep{Schady06,Oates06}. The peak of
GRB\,030418 might have been the afterglow onset. Setting $z = 1$, we get
$\Gamma_0 \sim 100$ using Eq.~(\ref{Gamma}).

\section{Conclusions}

The REM discovery of the afterglow onset has demonstrated once again the
richness and variety of physical processes occurring in the early afterglow
stages. The very fast response observations presented here provide crucial
information on the GRB fireball parameters, most importantly its initial
Lorentz factor. This is the first time that $\Gamma(t_{\rm peak})$ is directly
measured from the observations of a GRB. The measured $\Gamma_0$ value is well
within the range ($50 \la \Gamma_{0} \la 1000$) envisaged by the standard
fireball model \citep{Piran00,Guetta01,Alicia02,Meszaros}. It is also in
agreement with existing measured lower limits \citep{Lith01,zhang06}. Albeit
the values for $\Gamma_0$ are comparable for GRB\,060418 and GRB\,060607A, as
noted in the introduction prompt optical detections have shown very different
behaviours, so this value should not be taken as common to every GRB.

Using $\Gamma_0=400$ we can also derive other fundamental quantities
characterising the fireball of the two bursts. In particular, the
isotropic-equivalent baryonic load of the fireball is $M_{\rm fb} = E/(\Gamma_0
c^2) \approx 7\times 10^{-4}\,M_\odot$, and the deceleration radius is $R_{\rm
dec} \approx 2 c t_{\rm peak} [\Gamma(t_{\rm peak})]^2/(1+z) \approx
10^{17}$~cm. This is much larger than the scale of $\sim 10^{15}$~cm where the
internal shocks are believed to power the prompt emission
\citep{MesRees97,Rees94}, thus providing further evidence for a different
origin of the prompt and afterglow stages of the GRB.

\begin{acknowledgements}
We thank Yi-Zhong Fan, Zhi-Ping Jin, Massimo Della Valle and D. Alexander Kann
for discussion, and an anonymous referee for suggestions. The REM project has
been accomplished thanks to the support of the Italian Ministry of Education
(MURST/COFIN, P.I. G.C.). Part of the instrumentation has been partially
supported by ASI (ROSS  spectrograph, PI E.Pa.), by CNAA and by the
\textit{Swift} Project (supported by ASI); present maintenance and operation of
the REM telescope is supported by INAF (PI E.M.). We also acknowledge the
sponsorship of the Italian division of AMD who provided computers for the REM
observatory. REM is the result of a collaboration between a group of Italian
research institutes, coordinated by the Osservatorio Astronomico di Brera, the
Laboratoire Astroparticule et Cosmologie and the DSM/DAPNIA/Service
d'Astrophysique (France), the Dunsink Observatory and UCD (Ireland). S.D.V. and
D.M. are supported by SFI and IDA, respectively.
\end{acknowledgements}

\Online

\onltab{2}{
\begin{table}
\caption{Observation log for GRB\,060418. The time $t_0$ indicates the BAT
trigger time, 2006 April 18.12926 UT.\newline $^{*}$ \cite{Nyse06}.}
\centering\begin{tabular}{lcclcc}\hline
Mean time    & $t-t_0$  & Exp. time & Filter & Magnitude        \\
(UT)         & (s)      & (s)	    &	     &  		\\ \hline
Apr 18.12974 &      40  & 5	    & $z$    & $15.3 \pm 0.3^{*}$ \\
Apr 18.13371 &     479  & 100	    & $z'$   & $13.26 \pm 0.06$ \\
Apr 18.14215 &    1114  & 100	    & $z'$   & $14.70 \pm 0.09$ \\
Apr 18.15722 &    2416  & 100	    & $z'$   & $15.58 \pm 0.12$ \\ \hline
Apr 18.13264 &     292  & 100	    & $J$    & $12.33 \pm 0.05$ \\
Apr 18.13975 &     906  & 150	    & $J$    & $13.82 \pm 0.06$ \\
Apr 18.15323 &    2071  & 300	    & $J$    & $14.94 \pm 0.07$ \\ \hline
Apr 18.13014 &      76  & 10	    & $H$    & $11.98 \pm 0.16$ \\
Apr 18.13032 &      92  & 10	    & $H$    & $11.46 \pm 0.07$ \\
Apr 18.13050 &     107  & 10	    & $H$    & $11.00 \pm 0.19$ \\
Apr 18.13067 &     122  & 10	    & $H$    & $10.92 \pm 0.12$ \\
Apr 18.13085 &     137  & 10	    & $H$    & $11.01 \pm 0.02$ \\
Apr 18.13538 &     529  & 150	    & $H$    & $12.24 \pm 0.03$ \\
Apr 18.14527 &    1383  & 150	    & $H$    & $13.57 \pm 0.04$ \\
Apr 18.16008 &    2663  & 150	    & $H$    & $14.89 \pm 0.22$ \\
Apr 18.18578 &    4883  & 150	    & $H$    & $15.48 \pm 0.10$ \\
Apr 18.19649 &    5809  & 150	    & $H$    & $15.40 \pm 0.12$ \\
Apr 18.20720 &    6734  & 150	    & $H$    & $15.60 \pm 0.10$ \\
Apr 18.21791 &    7659  & 150	    & $H$    & $15.83 \pm 0.15$ \\ \hline
Apr 18.13121 &     168  & 10	    & $K'$   & $10.23 \pm 0.04$ \\
Apr 18.13139 &     184  & 10	    & $K'$   & $10.26 \pm 0.03$ \\
Apr 18.13157 &     200  & 10	    & $K'$   & $10.38 \pm 0.03$ \\
Apr 18.13175 &     215  & 10	    & $K'$   & $10.48 \pm 0.05$ \\
Apr 18.13192 &     230  & 10	    & $K'$   & $10.49 \pm 0.05$ \\
Apr 18.14652 &    1491  & 300	    & $K'$   & $13.14 \pm 0.09$ \\ \hline
\end{tabular}
\label{data1}
\end{table}
}

\onltab{3}{
\begin{table}
\caption{Observation log for GRB\,060607A. The time $t_0$ indicates the BAT
trigger time, 2006 June 7.21682 UT.}

\centering\begin{tabular}{lcclcc}\hline
Mean time   & $t-t_0$  & Exp. time & Filter & Magnitude      \\
(UT)        & (s)      & (s)	   &	    &		     \\ \hline
Jun 7.21767 &     73   &  10	   & $H$    & $14.60\pm0.20$ \\
Jun 7.21794 &     97   &  10	   & $H$    & $13.35\pm0.10$ \\
Jun 7.21811 &    111   &  10	   & $H$    & $12.91\pm0.07$ \\
Jun 7.21829 &    127   &  10	   & $H$    & $12.57\pm0.06$ \\
Jun 7.21848 &    143   &  10	   & $H$    & $12.33\pm0.05$ \\
Jun 7.21866 &    159   &  10	   & $H$    & $12.01\pm0.05$ \\
Jun 7.21883 &    174   &  10	   & $H$    & $12.18\pm0.05$ \\
Jun 7.21902 &    190   &  10	   & $H$    & $12.10\pm0.05$ \\
Jun 7.21919 &    205   &  10	   & $H$    & $12.07\pm0.05$ \\
Jun 7.21937 &    220   &  10	   & $H$    & $12.21\pm0.05$ \\
Jun 7.21955 &    236   &  10	   & $H$    & $12.21\pm0.05$ \\
Jun 7.21973 &    251   &  10	   & $H$    & $12.31\pm0.05$ \\
Jun 7.21991 &    267   &  10	   & $H$    & $12.40\pm0.05$ \\
Jun 7.22008 &    282   &  10	   & $H$    & $12.40\pm0.05$ \\
Jun 7.22027 &    298   &  10	   & $H$    & $12.55\pm0.06$ \\
Jun 7.22045 &    314   &  10	   & $H$    & $12.72\pm0.06$ \\
Jun 7.22062 &    328   &  10	   & $H$    & $12.59\pm0.06$ \\
Jun 7.22080 &    344   &  10	   & $H$    & $12.82\pm0.07$ \\
Jun 7.22098 &    359   &  10	   & $H$    & $12.84\pm0.07$ \\
Jun 7.22128 &    385   &  30	   & $H$    & $12.95\pm0.05$ \\
Jun 7.22169 &    421   &  30	   & $H$    & $12.96\pm0.05$ \\
Jun 7.22209 &    455   &  30	   & $H$    & $13.10\pm0.05$ \\
Jun 7.22251 &    492   &  30	   & $H$    & $13.23\pm0.05$ \\
Jun 7.22292 &    527   &  30	   & $H$    & $13.25\pm0.05$ \\
Jun 7.22400 &    620   & 150	   & $H$    & $13.40\pm0.04$ \\
Jun 7.22606 &    798   & 150	   & $H$    & $13.96\pm0.06$ \\
Jun 7.22812 &    976   & 150	   & $H$    & $14.32\pm0.07$ \\
Jun 7.23033 &   1167   & 150	   & $H$    & $14.94\pm0.08$ \\
Jun 7.23223 &   1331   & 150	   & $H$    & $14.63\pm0.07$ \\
Jun 7.23427 &   1508   & 150	   & $H$    & $14.69\pm0.08$ \\
Jun 7.23635 &   1687   & 150	   & $H$    & $14.55\pm0.07$ \\
Jun 7.24047 &   2043   & 150	   & $H$    & $14.49\pm0.07$ \\
Jun 7.24253 &   2221   & 150	   & $H$    & $14.68\pm0.08$ \\
Jun 7.24459 &   2399   & 150	   & $H$    & $15.05\pm0.09$ \\
Jun 7.24767 &   2665   & 300	   & $H$    & $15.46\pm0.08$ \\
Jun 7.25113 &   2964   & 300	   & $H$    & $15.73\pm0.10$ \\
Jun 7.25493 &   3293   & 300	   & $H$    & $15.58\pm0.09$ \\
Jun 7.25886 &   3632   & 150	   & $H$    & $15.48\pm0.13$ \\
Jun 7.37574 &  13730   & 6000	   & $H$    & $17.23\pm0.23$ \\ \hline
\end{tabular}

\label{data2}
\end{table}
}

\end{document}